\documentclass[aps,showpacs,floatfix,amssymb, twocolumn]{revtex4}

\usepackage{graphicx}
\begin{document}

\title{Magnetic order and spin excitations  in  the Kitaev--Heisenberg
model on the honeycomb lattice}

\author{ A.A. Vladimirov$^{a}$, D. Ihle$^{b}$ and  N. M. Plakida$^{a, c}$\footnote{E-mail: plakida@theor.jinr.ru} }
 \affiliation{ $^a$Joint Institute for Nuclear Research,
141980 Dubna, Russia}
 \affiliation{$^{b}$ Institut f\"ur Theoretische Physik,
 Universit\"at Leipzig,  D-04109, Leipzig, Germany }
\affiliation{$^{c}$  Max-Planck-Institut f\"{u}r Physik komplexer Systeme,
D-01187 Dresden,  Germany}

\date{\today}

\begin{abstract}
We consider the quasi-two-dimensional pseudo-spin-1/2 Kitaev - Heisenberg  model
proposed for A$_2$IrO$_3$ (A=Li, Na) compounds. The spin-wave excitation spectrum, the sublattice magnetization, and the transition temperatures  are calculated in the random phase approximation (RPA) for four different ordered phases, observed  in the parameter space of the model:  antiferomagnetic, stripe, ferromagnetic, and zigzag phases.  The N\'{e}el temperature and temperature dependence of the sublattice magnetization are compared with the experimental data on Na$_2$IrO$_3$.
\end{abstract}

\pacs{75.10.-b, 
 75.10.Jm, 
75.40.Cx 
 }

\maketitle

\section{Introduction}

Recent studies of transition-metal oxides have revealed an important role of the orbital degrees of freedom which  bring about  highly anisotropic spin interactions and complicated magnetic properties of these materials (for a review see ~\cite{Khaliullin05,Nussinov15}). Particularly, fascinating phase diagrams have been observed for  the $4d$ and $5d$ transition-metal oxides. In comparison with $3d$ compounds, they have weaker Coulomb correlations due to a delocalized character of $4d$ and $5d$ states,  but a much  stronger relativistic   spin-orbit coupling (SOC). The latter entangles the spin and orbital degrees of freedom,  and a new type of quantum state bands emerges  determined by the effective total angular moment $J_{eff}$. For the  iridium-based compounds with 5d electrons on $t_{2g}$ orbitals in the magnetic ion Ir$^{4+}$,  a strong SOC splits the broad $t_{2g}$ band into $J_{eff} = 3/2$ and  $J_{eff} = 1/2$ subbands.  Then even a weak Coulomb correlation  brings  about a Mott insulating state in the half-filled $J_{eff} = 1/2$ band~\cite{Kim08}. Based on the consideration of crystal-field splitting and  SOC for layered iridium compounds, an effective Heisenberg model for the pseudospins $S = 1/2$  with the compass-model anisotropy was proposed in Ref.~\cite{Jackeli09}:
\begin{equation}
H = J \sum_{\langle i j \rangle} {\bf S}_i \cdot {\bf S}_{j} + K  \sum_{\langle i j \rangle_{\gamma}} S_i^\gamma S_{j }^\gamma,
\label{KH1}
\end{equation}
where $J$ is the isotropic Heisenberg interaction for nearest neighbors (n.n.) $\langle i j \rangle$ and $K$ is the n.n. $\langle i j \rangle_{\gamma}$ bond-dependent Kitaev interaction~\cite{Kitaev06}. The superexchange interaction on the square lattices  in  A$_2$IrO$_3$ compounds (A = Na, Ba) with corner-sharing oxygen octahedra is predominantly of the isotropic\ Heisenberg type $J$,  while for the  honeycomb lattices  in A$_2$IrO$_3$ compounds (A = Li, Na) with edge-sharing oxygen octahedra the anisotropic Kitaev  interaction $K$ dominates.  The exact solution of the   Kitaev model~\cite{Kitaev06}  reveals a highly frustrated quantum spin-liquid phase with  peculiar  dynamics ~\cite{Knolle14,Knolle14a,Knolle15}. The inclusion of a finite  isotropic  Heisenberg interaction $J$ lifts  the degeneracy of the ground state, and  a rich phase diagram with competing long-range orders, such as the  ferromagnetic (FM), antiferromagnetic (AF), stripe and zigzag phases, emerges~\cite{Chaloupka10,Chaloupka13}.

 The parameters of the Kitaev-Heisenberg (KH) model (\ref{KH1}) for Na$_2$IrO$_3$ were  calculated  using the density functional theory \cite{Kim12,Kim13,Foyevtsova13,Yamaji14}, {\it ab initio} quantum chemistry calculations~\cite{Katukuri14,Nishimoto14}, and microscopic superexchange calculations~\cite{Sizyuk14}. As a general conclusion it was found that  for Na$_2$IrO$_3$ the n.n. Kitaev interaction is FM and   much stronger than the AF Heisenberg interaction, e.g., $K \simeq -17$~meV, $\, J \simeq 3$~meV~\cite{Katukuri14}.  For Li$_2$IrO$_3$ a strong dependence of the coupling constant on the parameters of Ir-O  bonds was found so that the n.n. Heisenberg  interaction $J$ has opposite signs for the two inequivalent Ir - Ir links:  $J \approx - 19$~meV  and  $J \approx 1$~meV for another link~\cite{Nishimoto14}. It was also found that the next n.n.  Heisenberg  and Kitaev interactions  are comparable to the n.n. contributions,  and they should be taken into account to describe the experimentally observed zigzag phase. In the absence of next n.n. interactions in the KH model (\ref{KH1}) the zigzag phase can be obtained only for AF Kitaev and FM Heisenberg interactions, e.g.,  $K \simeq 21$~meV, $ \, J \simeq - 4$~meV, as was proposed in Refs.~\cite{Chaloupka10,Chaloupka13}.  Depending on the  values of the  second ($J_2$) and third ($J_3$) neighbor Heisenberg interactions, a complicated phase diagram emerges with an incommensurate magnetic order in a large part of the diagram~\cite{Sizyuk14}.  An important role of the further-distant-neighbor interactions and of the  bond-depending  off-diagonal exchange interaction was also stressed in other publications (see Refs.~\cite{Kimchi11,Singh12,Rau14,Lou15}).

 The ground-state properties and excitation spectrum of the KH model have been studied by various methods, such as the Lanczos exact diagonalization  for finite clusters~\cite{Chaloupka10,Chaloupka13,Rau14}, pseudofermion renormalization group~\cite{Reuther11}, classical Monte Carlo simulation~\cite{Sizyuk14,Price12,Price13}, tensor variational approach\cite{Iregui14}, and the entanglement renormalization  ansatz~\cite{Lou15}. The spectrum of spin waves in the KH model  was calculated within  linear spin-wave theory (LSWT) in  the zigzag phase in Ref.~\cite{Chaloupka13}. In
 Refs.~\cite{You12,Hyart12,Okamoto13} doping effects on the phase diagram and emerging superconductivity in the extended KH model  were studied within a generalized $t$-$J$  model.

Most of  experimental studies  are devoted to Na$_2$IrO$_3$. Measurements of electrical resistivity, magnetization, magnetic susceptibility, and heat capacity of Na$_2$IrO$_3$ have shown a  phase  transition to the long-range AF order below $T_N =15$~K~\cite{Singh10}.  In Ref.~\cite{Liu11}, using resonant x-ray scattering, the AF phase transition was found at $T_N =13.3$~K, and the zigzag magnetic structure was proposed. A direct evidence of the  zigzag magnetic phase was obtained by neutron and x-ray diffraction investigations of Na$_2$IrO$_3$ single crystals below  $T_N =18$~K~\cite{Ye12}. In Ref.~\cite{Choi12} the spectrum of spin excitations in  Na$_2$IrO$_3$ was measured by inelastic  neutron scattering which confirmed the zigzag magnetic order. The spin-wave spectrum was observed below $5$~mev and was described within  LSWT for the Heisenberg model with the exchange interaction up to the third neighbors,  while the contribution from the Kitaev interaction was considered to be  small.  The  long-range magnetic order  below $T_N =15.3$~K in this study  was detected by  the  muon-spin rotation method. Magnetic excitations in Na$_2$IrO$_3$ were also investigated in Ref.~\cite{Gretarsson13}  using resonant inelastic x-ray scattering. Excitations  with  much higher energy of about $35$~meV were observed at the $\Gamma$ point in the Brillouin zone with the dispersion  consistent with the  calculation in  Ref.~\cite{Chaloupka13}. In Ref.~\cite{Comin12} optical and angle-resolved photoemission spectroscopy on Na$_2$IrO$_3$ revealed an insulating gap of 340~meV which can be explained  by suggesting a large Coulomb repulsion $U = 3$~eV in the Mott insulating state. In Ref.~\cite{Singh12} roughly the same
temperatures of the  magnetic phase transition, $T_N \approx 15$~K, in  A$_2$IrO$_3$  for A= Na and Li were reported using magnetic and heat capacity measurements  .

In the present paper we perform self-consistent calculations of the sublattice magnetization and the spin-wave excitation spectrum for the KH model  (\ref{KH1}) on the honeycomb lattice. We consider the full parameter space of the model, where four ordered phases are known to exist. To take into account the finite-temperature renormalization of the spectrum and to calculate the transition temperature $T_c$, we employ the equation of motion method for  Green functions (GFs)~\cite{Zubarev60} for spin $S = 1/2$  using the random phase approximation (RPA)~\cite{Tyablikov75}, as we have done for the compass-Heisenberg model on the square lattice  in Ref.~\cite{Vladimirov14}.

In Sec.~II we formulate the KH model  and derive equations for the matrix GF.  The magnetization and phase transition temperatures for all  four phases are considered in Sec.~III.  The results of spin-wave spectrum calculations and for the  phase diagram are presented in Sec.~IV.  They are compared with experiments on  A$_2$IrO$_3$ and other theoretical studies  of the KH  model. In Sec.~V the conclusion is given, and in the Appendix details of calculations are presented.

\section{Spin-excitation spectrum}

\subsection{Kitaev-Heisenberg model}

We consider the  KH model on the honeycomb lattice with the n.n.
distance $a_0$. The lattice is bipartite with two sublattices $A$ and $B$.
\begin{figure}
\includegraphics[width=0.35\textwidth]{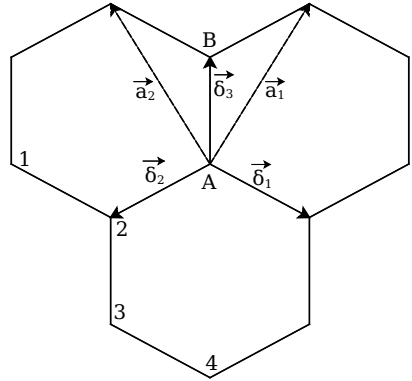}
\caption[]{Honeycomb lattice, where $\overrightarrow{\delta_1},\, \overrightarrow{\delta_2},\, \overrightarrow{\delta_3} $ are  the nearest-neighbour vectors (\ref{nn}), ${\bf a}_1$ and ${\bf a}_2$ are the lattice vectors. The  four sublattices in the zigzag and stripe phases are denoted by  the numbers  $1, 2, 3, 4$. }
 \label{fig1}
\end{figure}
Each lattice site on $A$ is connected to three n. n. sites belonging to $B$
by vectors $\overrightarrow{\delta_j}$, sites on $B$ sublattice are connected to $A$ by vectors $-\overrightarrow{\delta_j}$
 (see  Fig.~\ref{fig1}):
\begin{equation}
\overrightarrow{\delta_1} = \frac{a_0}{2}(\sqrt{3}, -1),\; \overrightarrow{\delta_2} = -  \frac{a_0}{2}(\sqrt{3}, 1), \; \overrightarrow{\delta_3} = a_0(0, 1).
\label{nn}
\end{equation}
The lattice vectors  are ${\bf a}_1 = \overrightarrow{\delta_3} - \overrightarrow{\delta_2} = ({a_0}/{2})(\sqrt{3}, 3)$ and ${\bf a}_2 = \overrightarrow{\delta_3} - \overrightarrow{\delta_1} = ({a_0}/{2})(-\sqrt{3}, 3)$,
the lattice constant is $a = |{\bf a}_1| = |{\bf a}_2| = \sqrt{3}a_0$. The reciprocal lattice is defined by the vectors
${\bf k}_1 = ({2\pi}/{3a_0}) (\sqrt{3}, 1)$  and $ {\bf k}_2 =({2\pi}/{3a_0}) (-\sqrt{3}, 1)$.

The KH model (\ref{KH1}) is convenient to write in a short notation as:
\begin{equation}
H = \sum_{i, m, \nu} J^{\nu}_m S^{\nu}_i S^{\nu}_{i + m} .
\label{H1}
\end{equation}
Here, $i$ goes over all sites of the  $A$ sublattice, and $i+m$ denotes
n.n. sites of  $i$, which belong to the $B$ sublattice,
${\bf r}_{i+m} = {\bf r}_i + \overrightarrow{\delta_m}$.
The exchange interaction $J^{\nu}_m$ depends on the spin component
index $\nu = x,y,z$ and the  bond number $m = 1,2,3$. In the particular
case of the KH model,  the exchange interaction reads as
$J^{\nu}_1 = (J+K_x, J, J), J^{\nu}_2 = (J, J+K_y, J),
J^{\nu}_3 = (J, J, J+K_z)$ where we can also consider an  anisotropic  Kitaev interaction, $K_x \neq K_y \neq K_z$.

In the general case we consider several sublattices for the model  (\ref{H1}) with the sublattice vectors ${\bf b}_j$, where $j$ is the sublattice index,
${\bf b}_1 \equiv 0$, ${\bf b}_2$ connects the first sublattice to the second one, etc.
Any vector connecting sites on the same sublattice is a combination
of the lattice vectors ${\bf a}_1$, ${\bf a}_2$. All ${\bf a}$ and ${\bf b}$ vectors are combinations of $\overrightarrow{\delta_i}$. To study the zigzag phase, we have  to consider the four sublattice representation as in Refs.~\cite{Choi12,Chaloupka10,Chaloupka13}.

Using the spin operators $S^\pm_i = S^x_i \pm i S^y_i, \,S^z_i $,  the Hamiltonian (\ref{H1}) can be written as:
\begin{eqnarray}
H = \sum_{i,j,k,l} && J^z_{i,j,k,l} S^z_{i,j} S^z_{k,l}
+ \frac{1}{2} J^+_{i,j,k,l} (S^+_{i,j} S^-_{k,l} + S^-_{i,j} S^+_{k,l}) \nonumber\\
&&+ \frac{1}{2} J^-_{i,j,k,l} (S^+_{i,j} S^+_{k,l} + S^-_{i,j} S^-_{k,l})   ,
\label{H2}
\end{eqnarray}
with $J^{\pm}_{i,j,k,l} = (1/2)(J^x_{i,j,k,l} \pm J^y_{i,j,k,l})$.
Here $i$,$k$ are lattice indexes, and $j$,$l$ are sublattice indexes.
The honeycomb lattice has two nonequivalent sites $A$ and $B$ per  unit cell. If we have more then two sublattices, we can define them in such a way, that lattice sites
with odd (even) sublattice indexes belong  to the sublattice $A (B)$.
Then  the interaction parameters $J^{\nu}$ for $\nu \in \{z, +, -\}$  have the form:
$J^{\nu}_{i,j,k,l} = \sum_{m} J^{\nu}_m \delta [{\bf a}_k + {\bf b}_l - {\bf a}_i - {\bf b}_j + (-1)^j
 \overrightarrow{\delta_m}]$, where  $\sum_{m} \delta [{\bf a}_k + {\bf b}_l - {\bf a}_i - {\bf b}_j + (-1)^j \overrightarrow{\delta_m}]$ is equal to unity if  the $(k,l)$ site is n.n.  of the $(i,j)$ site.
The  components $J^{\nu}_m$  are given by $J^z_1 = J^z_2 = J$, $J^z_3 = J + K_z$,
$J^+_1 = J + K_x/2$, $J^+_2 = J + K_y/2$, $J^+_3 = J$,
$J^-_1 = K_x/2$, $J^-_2 = -K_y/2$, $J^-_3 = 0$.

\subsection{Green function equations}

To calculate the spin-wave spectrum  of transverse spin
excitations, we introduce the matrix  retarded two-time commutator
GF~\cite{Zubarev60}:
\begin{eqnarray}
 &&\hat{G} (t-t') = -i \theta(t-t')
 \langle [
\hat{S}(t) \;  ,\; \hat{S}^{\dag}(t') ]\rangle
\nonumber\\
&&
= \int_{-\infty}^{+\infty}\,\frac{d\omega}{2\pi}\,e^{-i\omega(t-t')}
  \hat G(\omega),
     \label{r1}
\end{eqnarray}
where
\begin{eqnarray}
&&\hat G(\omega)= \left(
\begin{array}{cccc}
  \langle \! \langle  S^+_{i,j} | S^-_{i',j'} \rangle\! \rangle_\omega  \;
  \langle \! \langle  S^-_{i,j} | S^-_{i',j'} \rangle\! \rangle_\omega  \\
  \langle \! \langle  S^+_{i,j} | S^+_{i',j'} \rangle\! \rangle_\omega  \;
  \langle \! \langle  S^-_{i,j} | S^+_{i',j'} \rangle\! \rangle_\omega  \\
        \end{array}\right) .
\label{r1a}
\end{eqnarray}
Using the commutation relations for    spin operators, $\, [ {S_i^+} ,
{S_j^-} ]  = 2 {S_i^z} \, \delta_{i,j} , \quad
  [ {S_i^\pm}, {S_j^z}] =
  \mp {S_i^\pm}\, \delta_{i,j}  ,
\,$ we obtain  equations of motion for the GFs
\begin{eqnarray}
&&\omega \langle \! \langle  S_{i,j}^+ |
 S_{i',j'}^- \rangle\! \rangle_\omega =
2\langle S^z_{i,j} \rangle\, \delta_{i,i'}\delta_{j,j'}
 \nonumber \\
&& - \sum_{k,l} \, J^z_{i,j,k,l} \langle \! \langle \,
 S^+_{i,j} S^z_{k,l} | S_{i',j'}^- \rangle\! \rangle_\omega
 \nonumber \\
&& + \sum_{k,l} J^+_{i,j,k,l}
  \langle \!\langle \,
  S^z_{i,j} S^+_{k,l} | S_{i',j'}^-  \rangle\! \rangle_\omega
 \nonumber \\
&& + \sum_{k,l} \, J^-_{i,j,k,l} \langle \!\langle \, S^z_{i,j} S^-_{k,l} |
S_{i',j'}^-  \rangle\! \rangle_\omega,
\label{r2}\\
&&\omega \langle \! \langle  S_{i,j}^- |
 S_{i',j'}^+ \rangle\! \rangle_\omega =- 2\langle S^z_{i,j} \rangle\, \delta_{i,i'}\delta_{j,j'}
 \nonumber \\
&& + \sum_{k,l} \, J^z_{i,j,k,l} \langle \! \langle \,
 S^-_{i,j} S^z_{k,l} | S_{i',j'}^+ \rangle\! \rangle_\omega
 \nonumber \\
&& - \sum_{k,l} J^+_{i,j,k,l}
  \langle \!\langle \,
  S^z_{i,j} S^-_{k,l} | S_{i',j'}^+  \rangle\! \rangle_\omega
 \nonumber \\
&&- \sum_{k,l} \, J^-_{i,j,k,l} \langle \!\langle \, S^z_{i,j} S^+_{k,l} |
S_{i',j'}^+  \rangle\! \rangle_\omega  .
 \label{r3}
\end{eqnarray}
In the RPA~\cite{Tyablikov75}  for all GFs we use the following
approximation:
\begin{eqnarray}
\langle \! \langle S_{i,j}^z S_{k,l}^\alpha | S_{i',j'}^\beta \rangle\!
\rangle_\omega& =
s(j)\sigma\, \langle \! \langle S_{k,l}^\alpha  | S_{i',j'}^\beta
\rangle\! \rangle_\omega,
 \label{r4}
\end{eqnarray}
where $\sigma$ is the absolute value  of the order parameter while   $s(j) = \pm 1$ is the sublattice-dependent
sign  of the order parameter. By choosing $s(j)$  we can describe different phases in our model.

Using the momentum   representation  with respect to the lattice index $i$,
\begin{eqnarray}
  S_{i,j}^{\pm} = \sqrt{\frac{1}{N}}
  \sum_{\bf q}\,   S^{\pm}_{{\bf q}, j}\,
    {\rm e}^{{\pm}i{\bf q} ({\bf a}_i + {\bf b}_j)},
     \nonumber\\
  J^{\nu}_{{\bf q},j,l} = \frac{1}{N} \sum_{i,k} J^{\nu}_{i,j,k,l}
  {\rm e}^{i{\bf q}({\bf a}_k + {\bf b}_l - {\bf a}_i - {\bf b}_j)},
  \label{r5}
\end{eqnarray}
where $N$ is number of sites per sublattice, and introducing the nonation:
\begin{equation}
\gamma^z_j = \sum_{l} \, s (l) J^z_{0,j,l}, \;\;\;\;
\gamma^{\pm}_{{\bf q}, j, l} = s(j) J^{\pm}_{{\bf q},j,l},
\label{gamma}
\end{equation}
 Eqs.~(\ref{r2}) and (\ref{r3}) for the GFs in the RPA (\ref{r4})  can be written as
\begin{eqnarray}
&&\omega \langle \! \langle  S_{{\bf q}, j}^+ |
 S_{{\bf q}, j}^- \rangle\! \rangle_\omega =
 2\sigma s(j)
- \sigma \gamma^z_j \langle \! \langle \, S^+_{{\bf q}, j} | S_{{\bf q}, j}^- \rangle\! \rangle_\omega
\label{GF_eq1}\\
&&+ \sigma \sum_{l} \gamma^+_{{\bf q}, j, l}
  \langle \!\langle \, S^+_{{\bf q}, l} | S_{{\bf q}, j}^-  \rangle\! \rangle_\omega
+ \sigma \sum_{l} \gamma^-_{{\bf q}, j, l}
  \langle \!\langle \, S^-_{{-\bf q}, l} | S_{{\bf q}, j}^-  \rangle\! \rangle_\omega,
\nonumber\\
&&\omega \langle \! \langle  S_{{\bf q}, j}^- |
 S_{{\bf q}, j}^+ \rangle\! \rangle_\omega =
 -2\sigma s(j)
+ \sigma \gamma^z_j \langle \! \langle \, S^-_{{\bf q}, j} | S_{{\bf q}, j}^+ \rangle\! \rangle_\omega
\label{GF_eq2}\\
&&- \sigma \sum_{l} \gamma^+_{{\bf q}, j, l}
  \langle \!\langle \, S^-_{{\bf q}, l} | S_{{\bf q}, j}^+  \rangle\! \rangle_\omega
- \sigma \sum_{l} \gamma^-_{{\bf q}, j, l}
  \langle \!\langle \, S^+_{{\bf q}, l} | S_{{-\bf q}, j}^+ \rangle\! \rangle_\omega.
\nonumber
\end{eqnarray}
The system of $\, 2n \,$ equations for $n$ sublattices  can be written in the matrix form:
\begin{equation}
\omega \langle \!\langle \, \mathbb{S} | \mathbb{S}^\dagger \rangle\! \rangle_{{\bf q},\omega} =
\hat{\sigma} + \sigma \hat{V}(\bf q) \langle \!\langle \, \mathbb{S} | \mathbb{S}^\dagger \rangle\! \rangle_{{\bf q},\omega},
\label{m_eq}
\end{equation}
where $\mathbb{S} = [S^+_1, S^-_1, S^+_2, S^-_2...]$, $\hat{\sigma} = [2\sigma s(1), -2\sigma s(1), 2\sigma s(2), -2\sigma s(2)...]$,
and $\hat{V}$ is the matrix of the $\gamma$ coefficients (\ref{gamma}).
This system of equations has  the solution
\begin{equation}
\langle \!\langle \, \mathbb{S} | \mathbb{S}^\dagger \rangle\! \rangle_{{\bf q},\omega} =
[\, \omega \hat{I} - \sigma \hat{V}(\bf q)]^{-1}\hat{\sigma},
\label{sw_eq}
\end{equation}
where $\hat{I}$ is the unity matrix. The spectrum of spin excitations is given by the eigenvalues of the matrix $\sigma \hat{V}$.

\section{Magnetic order}

To calculate the sublattice magnetization $\sigma = \langle S_i^z
\rangle $ in RPA, we   use  the kinematic relation $\, S_i^z =
(1/2) - S_i^- S_i^+\,$ for spin $S=1/2$ which results in  the
self-consistent equation
 \begin{eqnarray}
 \sigma = \langle S_i^z\rangle
  = \frac{1}{2} - \frac{1}{N}
 \sum_{\bf q}\, \langle S^-_{\bf q} S^+_{\bf q}\rangle .
 \label{m6}
\end{eqnarray}
The correlation function in Eq.~(\ref{m6}) is calculated from the GF (\ref{sw_eq}) using the spectral representation,
\begin{equation}
\langle S^-_{\bf q}  S^+_{\bf q}\rangle
 = 2\sigma\,\sum_{i} I_{i}({\bf q}) \, N[\omega_{i} ({\bf q})]\,,
 \label{m5}
\end{equation}
where $N(\omega)= [\exp (\omega/T) - 1]^{-1}$, $\omega_{i}({\bf q}) = \sigma \varepsilon_{i}({\bf q})$,  $\varepsilon_{i}({\bf q})$ are eigenvalues of $\hat{V}(\bf q)$, and
\begin{eqnarray}
I_{i}({\bf q}) = \frac{a^{11}_{\bf q}[\varepsilon_{i}({\bf q})]}
{\prod_{j \neq i} [\varepsilon_{i}({\bf q}) - \varepsilon_{j}({\bf q})]}.
 \label{m5a}
\end{eqnarray}
Here $a^{11}_{\bf q}[\varepsilon_{i}({\bf q})]$ is the  $(1,1)$ first minor of the $ [\varepsilon
\hat{I} - \hat{V}(\bf q)]$ matrix.

By taking the limit $\sigma \rightarrow 0$ we can also obtain an
equation for the N\'{e}el temperature:
\begin{equation}
\frac{1}{T_N} = \frac{4}{N} \sum_{\bf q}\, \sum_{i} \frac{ I_{i}({\bf q})}  {\varepsilon_{i} ({\bf q})}.
 \label{m6a}
\end{equation}
The sum over ${\bf q}$ in this equation will diverge in the two-dimensional (2D) case if the spin excitation spectrum has no gaps, i.e., $\varepsilon_{i} ({\bf Q}) = 0$  at some momentum ${\bf Q}$. In this case,  in order to obtain a finite transition
temperature, we either should  consider the  3D case  introducing an inter-plane coupling $J_{\bot}$ (either  FM or AF) or add a small anisotropy to the  Kitaev interaction, e.g.,  $K_z > K_y = K_x$.  This opens  a gap at this wave vector,  as discussed  in the next section.

\section{Results and discussion}

\begin{figure}
\includegraphics[width=0.42\textwidth]{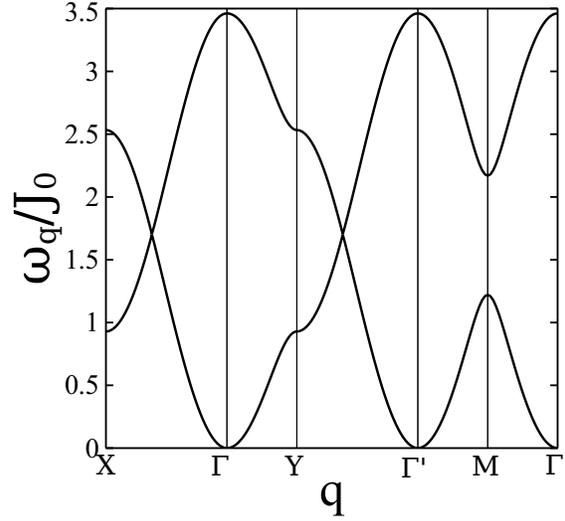}
\caption[]{Spin-wave spectrum for the  FM phase at $\phi = 200^{\circ}$, where  $J = J_0 \, \cos \phi$, $K = 2 \, J_0 \,\sin \phi$ .}
 \label{fig2}
\end{figure}
\begin{figure}
\includegraphics[width=0.42\textwidth]{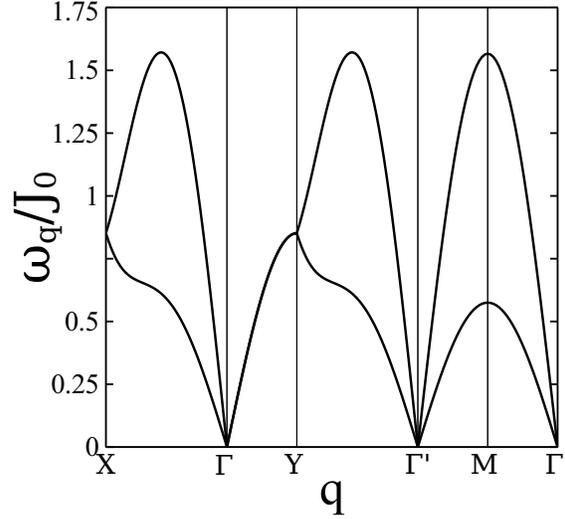}
\caption[]{Spin-wave spectrum for the AF phase at $\phi = 50^{\circ}$.}
 \label{fig3}
\end{figure}
\begin{figure}
\includegraphics[width=0.42\textwidth]{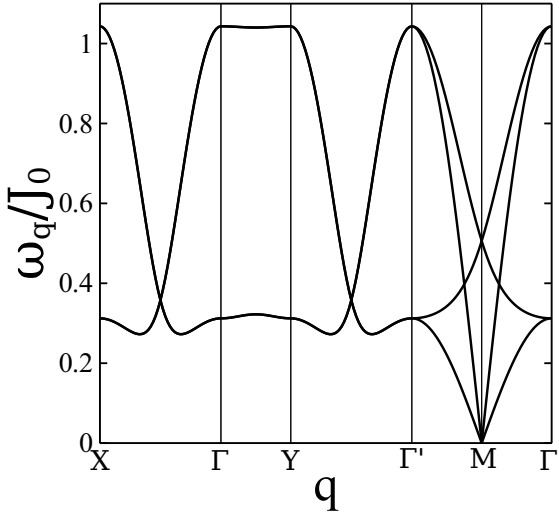}
\caption[]{Spin-wave spectrum for the zigzag phase at $\phi = 110.85^{\circ}$.}
 \label{fig4}
\end{figure}
\begin{figure}
\includegraphics[width=0.42\textwidth]{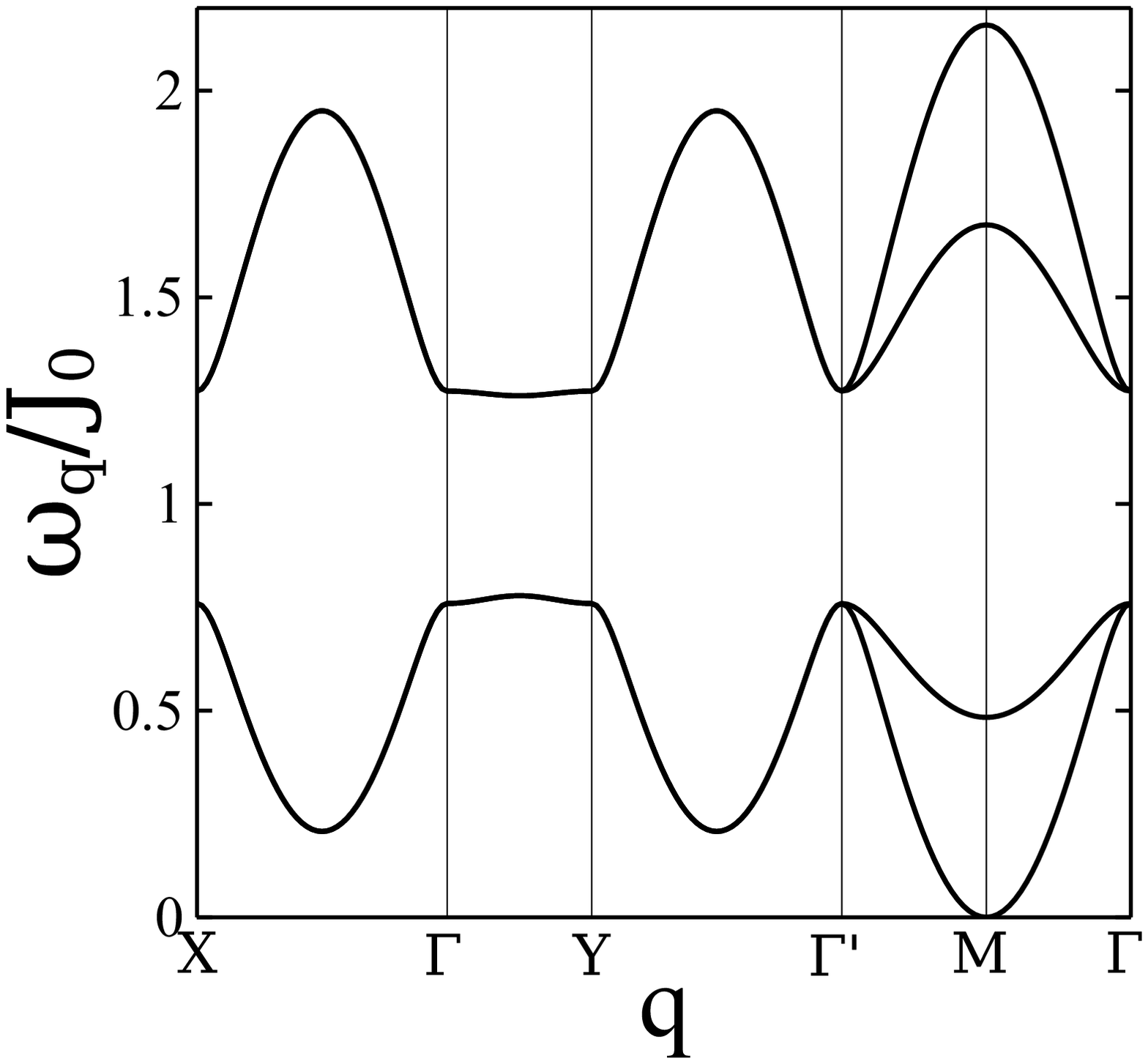}
\caption[]{Spin-wave spectrum for the atripe phase at $\phi = 300^{\circ}$.}
 \label{fig5}
\end{figure}
\begin{figure}
\includegraphics[width=0.42\textwidth]{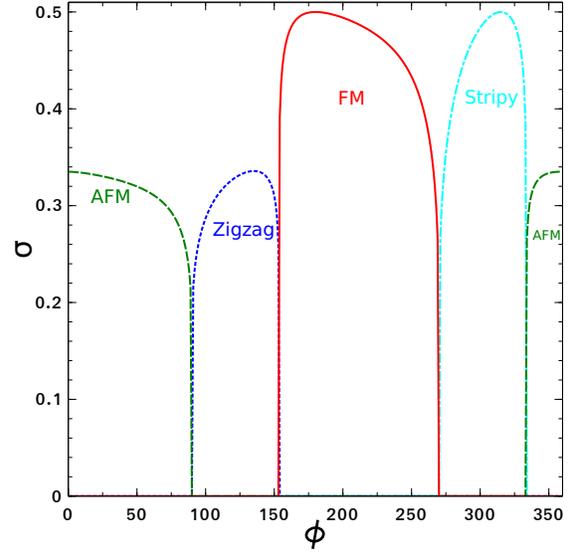}
\caption[]{(Color online) Magnetization for different phases at zero temperature versus phase angle  $\phi$.}
 \label{fig6}
\end{figure}
\begin{figure}
\includegraphics[width=0.42\textwidth]{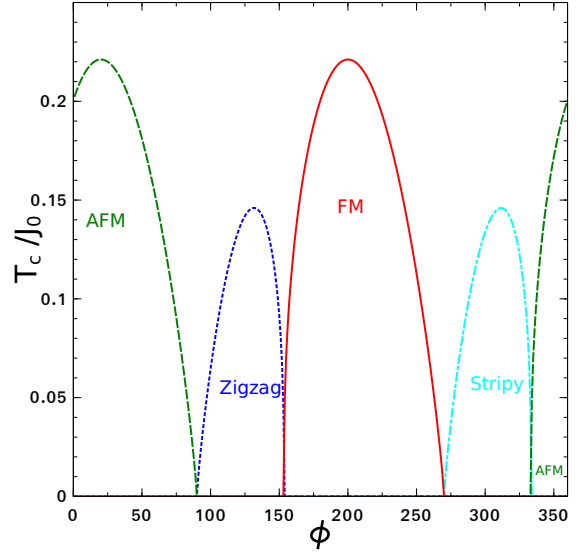}
\caption[]{(Color online)  Transition temperature for different phases versus phase  angle  $\phi$ at $J_\bot = - 0.0018 \, J_0$.}
 \label{fig7}
\end{figure}
\begin{figure}
\includegraphics[width=0.42\textwidth]{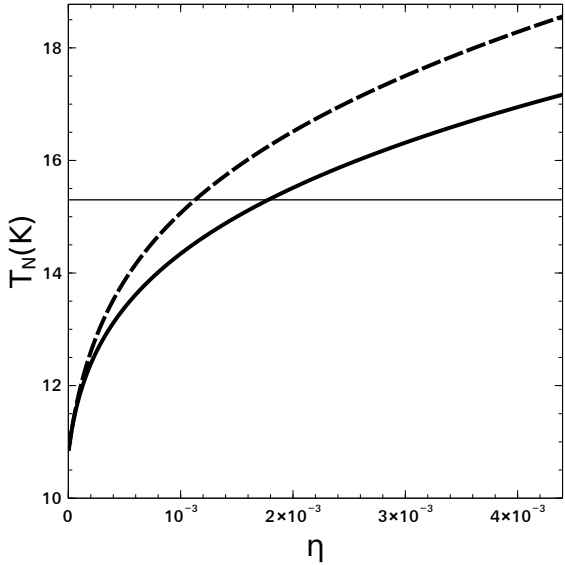}
\caption[]{Ne\'{e}l temperature $T_N$ as a function of the interplane coupling $J_\bot = -\eta J_0$ (solid line) and anisotropy
$K_z = (1 + \eta) K$, $K_x = K_y = K$ (dashed line), with
$J = -4$ meV, $K = 21$ meV, compared with the  experimental Ne\'{e}l temperature of Na$_2$IrO$_3$  (thin line) from \cite{Choi12}.}
 \label{fig8}
\end{figure}
\begin{figure}
\includegraphics[width=0.42\textwidth]{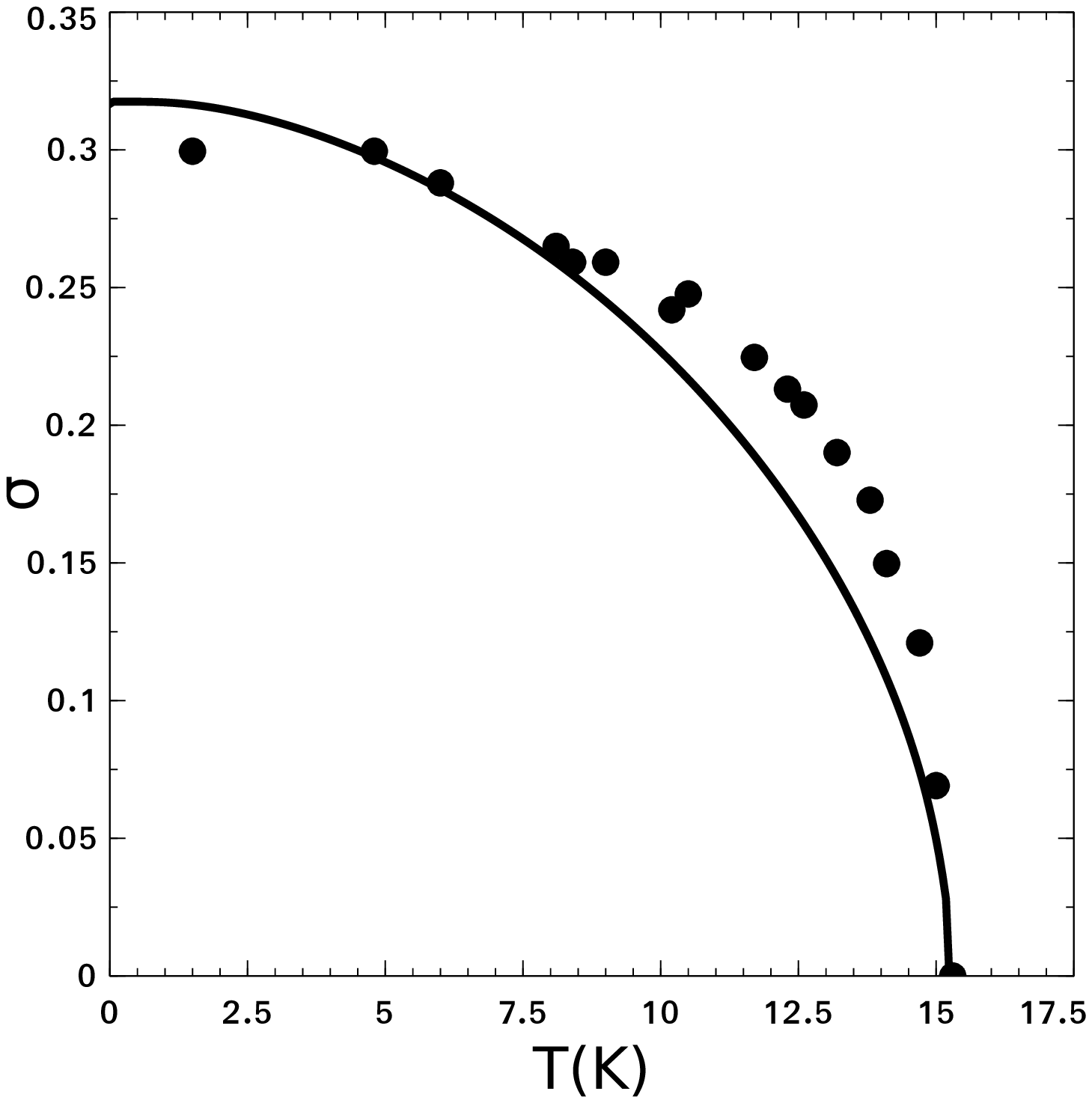}
\caption[]{Sublattice magnetization $\sigma$ as a function of temperature in the zigzag phase for $J = -4$ meV, $K = 21$ meV,
$J_\bot = - 0.02$ meV (solid line), compared with experimental data on Na$_2$IrO$_3$ from \cite{Choi12} in arbitrary units (circles).}
 \label{fig9}
\end{figure}
In this section we calculate the spin-wave spectrum  $\omega_{i}({\bf q}) $ for the AF, FM, zigzag, and stripe phases by solving  Eq.~(\ref{sw_eq}) and determine self-consistently the sublattice magnetization $\sigma = \langle S_i^z
\rangle$ in these phases using  Eq.~(\ref{m6}).
In the equations for the spin-wave spectra  we introduce the
short notations: $c_x = \cos(\frac{\sqrt{3}}{2} a_0 q_x)$,
$s_x =\sin(\frac{\sqrt{3}}{2} a_0 q_x)$,
$c_y = \cos(\frac{3}{2} a_0 q_y)$, and $s_y = \sin(\frac{3}{2} a_0 q_y)$.
In the AF phase we get:
\begin{equation}
\varepsilon^2_{\pm}({\bf q}) = A^2 + |B_{\bf q}|^2 - |C_{\bf q}|^2
\pm 2\sqrt{A^2|B_{\bf q}|^2 - [{\rm Im}(B_{\bf q}C^*_{\bf q})]^2},
\label{afm_spec}
\end{equation}
In the FM phase we have:
\begin{equation}
\varepsilon^2_{\pm}({\bf q}) = A^2 - |B_{\bf q}|^2 + |C_{\bf q}|^2
\pm 2\sqrt{A^2|C_{\bf q}|^2 - [{\rm Im}(B_{\bf q}C^*_{\bf q})]^2}.
\label{fm_spec}
\end{equation}
Here
\begin{eqnarray}
A &=& 3J + K_z, \nonumber\\
|B_{\bf q}|^2 &=& K_{+}^2 - K_x K_y c_x^2, \nonumber\\
|C_{\bf q}|^2 &=& [(2J+ K_{+}) c_x + J c_y]^2 \nonumber\\
&+&(J s_y + K_{-} s_x)^2, \nonumber\\
{\rm Im}(B_{\bf q} C^*_{\bf q}) &=& K_{+}
s_x [(2J+K_{+}) c_x + J c_y] \nonumber\\
&-& K_{-} c_x (K_{-} s_x - J s_y),
\label{fm_afm_e}
\end{eqnarray}
where $\,K_{\pm} = (K_x \pm K_y)/2\,$.  For the
zigzag phase we obtain Eq.~(\ref{zz_spec}) which for the  isotropic intaraction $K_x = K_y = K_z = K$  can be simplified to get:
\begin{eqnarray}
 && \varepsilon_{1,2}^2({\bf q}) = (4J^2 + 4KJ + 2K^2)c_x^2 - 2KJ (1+s_x s_y)
\nonumber\\
 && \pm 4| (J + K/2)\,c_x| [(K - J)^2 - (J s_y + K s_x)^2]^{1/2},
  \label{zz_spec_s}
\end{eqnarray}
and $\,\varepsilon_{3,4}(q_x, q_y) =  \varepsilon_{1,2}( - q_x, q_y)\,$.
For the stripe phase we have Eq.~(\ref{str_spec}) which for the isotropic interaction
can be simplified as:
{
\small
\begin{eqnarray}
&&\varepsilon_{1,2}^2({\bf q}) = 2K^2 s_x^2 - (2J^2 + 4KJ) c_x^2 - 2KJ(1-s_y s_x)  \pm M,  \nonumber\\
&&M^2 = (K - J)^2 (4J^2 + K^2)
 \nonumber\\
&&-4c_x^2[(2 K J + K^2)^2 s_x^2 - 4(2 J^2 + K J)^2 s_y^2]
 \nonumber\\
&&+4J K s_y s_x[2(K - J)^2 - 8(J + K/2)^2 c_x^2],
\label{str_spec_s}
\end{eqnarray} }
and $\,\varepsilon_{3,4}(q_x, q_y) =  \varepsilon_{1,2}( - q_x, q_y)\,$.
In the general case $K_x \neq K_y \neq K_z\,$,  Eqs.~ (\ref{zz_spec}), (\ref{zz_str_e}), and (\ref{str_spec}) should be used.

To compare our RPA results with the  exact diagonalization data from \cite{Chaloupka13}, we introduce the same notation for the  model parameters  $J = J_0 \, \cos \phi, \;\;\; K_x = K_y = K_z = 2 J_0 \, \sin \phi$ with the same energy unit $J_0 = \sqrt{(K/2)^2 + J^2}\, $ (in the model (\ref{KH1}) we use the parameter $K$ twice as large as in~\cite{Chaloupka13}). For  $J = -4$ meV and $K = 21$ meV  suggested for Na$_2$IrO$_3$ in \cite{Chaloupka13}, the energy unit equals to  $J_0 = 11.24$ meV.
\par
Spin-wave spectra for different values of $\phi$ corresponding to the four ordered phases are shown  in Figs.~\ref{fig2}--\ref{fig5} along the symmetry directions $X(-1, 0) \rightarrow \Gamma(0, 0) \rightarrow Y(0, 1) \rightarrow \Gamma'(1, 1)\rightarrow M(1/2,1/2)   \rightarrow \Gamma  $. The spectrum has a quadratic dispersion  $\varepsilon_{i}({\bf q}) \propto q^2$ at small $q$ close to the $\Gamma, \Gamma'$-points for the FM phase and close to the $M$-point for the stripe phase. A linear dispersion  $\varepsilon_{i}({\bf q}) \propto q$ at small $q$ is observed for the AF phase  close  to $\Gamma, \Gamma'$-points and for the zigzag phase close to the $M$-point. The spin-wave spectrum for the zigzag phase $\omega^2_{i}({\bf q}) = \sigma^2 \varepsilon^2_{i}({\bf q})$,  where $\varepsilon^2_{i}({\bf q})$ is given by Eq.~(\ref{zz_spec_s}),  coincides with the  LSWT spectrum obtained in Ref.~\cite{Chaloupka13} and Ref.~\cite{Choi12} if we substitute  $\sigma = S = 1/2$. In our theory the excitation energy is lower in the AF and zigzag phases, since the  magnetization obtained in  RPA, $\sigma \simeq 0.32$, is smaller than $\sigma = 1/2$ in   LSWT due to zero-point fluctuations.
 In Ref.~\cite{Choi12} only a lower part of the spectrum was observed, below 5~meV, while in Ref.~\cite{Gretarsson13} a spin-excitation energy of about $35$~meV was found at the $\Gamma$ point with the dispersion similar to  the calculations in Ref.~\cite{Chaloupka13}: a large dispersion along the $\Gamma \rightarrow X$ direction and a much weaker one along the $\Gamma \rightarrow Y$ direction (as in Fig.~\ref{fig4}).   However, the excitation  energy  is much higher than $\omega(\Gamma) \simeq 19$~meV in Ref.~\cite{Chaloupka13} and our result $\omega(\Gamma) \simeq 1.04\, J_0 = 11.7$~meV. To fit the experimental value to our result  we should use a much larger energy unit $J_0 \simeq 34$~meV.

In  Fig.~\ref{fig6} the dependence of the sublattice magnetization $\sigma$ at zero
temperature as a function of $\phi$ is shown for different phases. The  positions of the four ordered phases are consistent with the phase diagram in \cite{Chaloupka13}. However, in RPA we cannot obtain spin-liquid phases in regions of small $J$, we have only two points $\phi = \pi / 2$ and $\phi = 3 \pi / 2$ where long-range order disappears.
As expected, the  points $(J, K)$ and $(-J, K + 2J)$ on the phase diagram have the same $\sigma$ and $T_c$. We have a fully polarized  ground state ($\sigma = 0.5$) at $\phi = \pi$ and $\phi = ({7 \, \pi}/{4})$, as has been also analytically shown in Ref.~\cite{Khaliullin05}.  The transitions from the zigzag to the  FM phase and from the atripe to the AF phase are rather sharp which can be considered as a first-order transition. The other two transitions are very smooth like at a  second-order transition.

To obtain a finite transition temperature, an interplane coupling $J_{\bot}$  or a  small anisotropy, $K_z = (1 +\eta) K $,  should be introduced. In Fig.~\ref{fig7} the  transition temperature is shown for all phases when  the small interplane coupling $J_\bot = - 0.0018 \, J_0$  is taken. The general dependence of the  Ne\'{e}l temperature as a function of $J_{\bot}$ and the anisotropy parameter $\eta$   is plotted in Fig.~\ref{fig8}.  So, the experimental value of $T_N = 15.3$~K \cite{Choi12} can be obtained either by using $J_{\bot} = - 0.0018 \, J_0$ or $\eta = 1.1 \times 10^{-3}$. In Fig.~\ref{fig9} the sublattice magnetization in the zigzag phase as a function of temperature is depicted. It has a similar temperature dependence as the experimental curve for Na$_2$IrO$_3$~\cite{Choi12} given   in arbitrary units.

\section{Conclusion}

In the present paper we have calculated the zero-temperature magnetization, transition temperature, and the temperature-dependent spin-wave spectrum for four phases of the KH model excluding the spin-liquid phase which cannot be obtained  in RPA due to the lack of  long-range order.  We have used the model (\ref{KH1}) with n.n. interaction parameters suggested for Na$_2$IrO$_3$  in Ref.~\cite{Chaloupka13}, $J = -4$ meV, $K = 21$ meV, which enabled us to obtain the phase diagram similar to Ref.~\cite{Chaloupka13} except the spin-liquid phase. However, as discussed in the Introduction,  further studies have shown that the n.n. Heisenberg interaction is AF, $J > 0$, while the Kitaev interaction is FM, $K <0 $. To explain the experimentally observed zigzag phase in this case, further-distant-neighbor interactions should be taken into account. In particular, in  Ref.~\cite{Sizyuk14} a minimal super-exchange model was proposed, where in addition to the n.n. interactions  further-neighbor Heisenberg interactions $J_2 <0, J_3 > 0$ and the Kitaev interaction $K_2 = -2J_2 >0$ are included.  In our theory these  distant-neighbor interactions can be also included in the equations of motion for the GFs (\ref{r2}),  (\ref{r3}) which results in a more complicated system of equations for the spin-wave spectrum and the corresponding  equation for the magnetization  (\ref{m6}). The results of these more extended calculations will be published elsewhere.

In the present study we have considered  four phases with long-rang order with a definite order parameter. To investigate  the thermodynamic properties, such as the spin susceptibility and heat capacity,  the paramagnetic phase should be considered.  For this we can use the generalized mean-field approximation to obtain a self-consistent system of equations for the GFs and correlation functions, as has been performed for the  compass-Heisenberg model on the square lattice in Ref.~\cite{Vladimirov15}.


\acknowledgments
\indent
One of the authors (N.P.) thanks the Directorate of the MPIPKS
for the hospitality extended to him during his stay at the Institute.
 Partial financial support by the Heisenberg-Landau program of JINR is
acknowledged.\\

\appendix
\section{Technical Details}

This section contains some details how equations for the GFs were obtained
for different phases using our general RPA results from Sec.~II.B. To do this, we transform the model Hamiltonian (\ref{H1}) into Eq.~(\ref{H2}) and substitute the result into the equations for the GFs (\ref{GF_eq1}), (\ref{GF_eq2}). Then we calculate   the eigenvalues and the first minor of the matrix $[\varepsilon \hat{I} - \hat{V} ]$ in Eq.~(\ref{sw_eq}) and use them to calculate the magnetization self-consistently.

\subsection{ AF and FM phases}

For the AF phase we have:
$J^{\nu}_{0,j,k,l} = \sum_{m} J^{\nu}_m \delta ({\bf a}_k + {\bf b}_l - {\bf b}_j + (-1)^j \overrightarrow{\delta_m})$
with the sublattice index $j = 1,2$  and s(l) = -s(j).  Since $l \neq j$ we have only one $l$ for a given $j$, so we obtain
${\bf a}_k + {\bf b}_l - {\bf b}_j = -(-1)^j \overrightarrow{\delta_m}$ and
\begin{eqnarray}
\gamma^z_j &=& - s (j) \sum_{m} J^{z}_m, \nonumber\\
\gamma^{\pm}_{{\bf q}, j} &=&  s(j) \sum_{m} J^{\pm}_m \exp [-i{\bf q}((-1)^j \overrightarrow{\delta_m})],
\end{eqnarray}
or in the matrix form (\ref{m_eq}):
\begin{equation}
\hat{V} = \pmatrix{A&0&C_{\bf q}&B_{\bf q}\cr 0&-A&-B_{\bf q}&-C_{\bf q}\cr -C_{-\bf q}&-B_{-\bf q}&-A&0
 \cr B_{-\bf q}&C_{-\bf q}&0&A\cr}.
\end{equation}
The eigenvalues of $\hat{V}$  are given by Eq.~(\ref{afm_spec}),
and for the first minor of the matrix $(\varepsilon \hat{I} - \hat{V})$  we have:
\begin{eqnarray}
a^{11}_{\bf q}(\varepsilon) = \varepsilon^3 + A\varepsilon^2 -
\varepsilon(A^2 + |B_{\bf q}|^2
 \nonumber\\
- |C_{\bf q}|^2) - A (A^2 - |C_{\bf q}|^2 - |B_{\bf q}|^2),
\end{eqnarray}
with
\begin{equation}
A = \sum_{j} J^z_j, \; B_{\bf q} = \sum_{j} J^-_j e^{i{\bf q \delta}_j}, \;
C_{\bf q} = \sum_{j}J^{+}_{j} e^{i{\bf q \delta}_j }.
 \label{ABC}
\end{equation}
By substituting here the exchange interaction components $J^{\nu}_m$  we get Eq.~(\ref{fm_afm_e}).

To obtain equations for the FM phase, we use the same function
$J^{\nu}_{0,j,k,l}$ as in the AF phase, but
 $ s(l) = s(j) = 1$, so that
\begin{equation}
\gamma^z_j =  \sum_{m} J^{z}_m, \; \; \; \;
\gamma^{\pm}_{{\bf q}, j} = \sum_{m} J^{\pm}_m \exp \{-i{\bf q}[(-1)^j \overrightarrow{\delta_m}]\},
\end{equation}
which yields:
\begin{equation}
  \hat{V} = \pmatrix{-A&0&C_{\bf q}&B_{\bf q}\cr 0&A&-B_{\bf q}&-C_{\bf q}\cr C_{-\bf q}&B_{-\bf q}&-A&0
 \cr -B_{-\bf q}&-C_{-\bf q}&0&A\cr},
\end{equation}
with the eigenvalues (\ref{fm_spec}) and
\begin{eqnarray}
a^{11}_{\bf q}(\varepsilon) = \varepsilon^3 - A\varepsilon^2 -
\varepsilon(A^2 - |B_{\bf q}|^2
\nonumber\\
 + |C_{\bf q}|^2) + A (A^2 - |C_{\bf q}|^2 - |B_{\bf q}|^2),
\end{eqnarray}
with the  same functions $A,B_{\bf q},C_{\bf q}$ as in the AF phase.
We still have two branches in the FM phase due to  two sites per   unit cell of the honeycomb lattice.

\subsection{Zigzag phase}

In the zigzag phase we have four sublattices $j = 1,2,3,4$ (see Fig.~\ref{fig1}) with the following order parameter signs:
$s(1) = s(2) = 1$, $s(3) = s(4) = -1$.
Now we substitute the  KH exchange interaction $J^{\nu}_{m}$
corresponding to the $\overrightarrow{\delta_m}$ bond
and obtain:
\begin{eqnarray}
\gamma^z_1 &= &\gamma^z_2 = -\gamma^z_3 =- \gamma^z_4 = -A,  \nonumber\\
A &= &  J^z_3 - J^z_1 - J^z_2.
\label{A8}
\end{eqnarray}
For $\gamma^{\pm}_{{\bf q}, j, l }$ (\ref{gamma})  we have
\begin{eqnarray}
&&\gamma^{\pm}_{{\bf q}, 1, 4} =  \gamma^{\pm}_{{-\bf q}, 2, 3} = -\gamma^{\pm}_{{\bf q}, 3, 2}= - \gamma^{\pm}_{{-\bf q}, 4, 1}  \equiv B^{\pm}_{\bf q},
\nonumber\\
&&\gamma^{\pm}_{{\bf q}, 1, 2} = \gamma^{\pm}_{-{\bf q}, 2, 1} = -\gamma^{\pm}_{{\bf q}, 3, 4} = - \gamma^{\pm}_{-{\bf q}, 4, 3} \equiv C^{\pm}_{\bf q},
 \label{A9}
\end{eqnarray}
where
\begin{eqnarray}
 B^{\pm}_{\bf q}& = &J^{\pm}_3 \exp (i{\bf q}\overrightarrow{\delta_3}) ,
 \nonumber\\
C^{\pm}_{\bf q}& =&J^{\pm}_1 \exp (i{\bf q}\overrightarrow{\delta_1}) + J^{\pm}_2 \exp (i{\bf q}\overrightarrow{\delta_2}) .
\label{A9a}
\end{eqnarray}
Now we substitute these $\gamma$ functions into Eqs.~(\ref{GF_eq1}), (\ref{GF_eq2})
introducing shorter notations:
$B^+_{\bf q} \equiv B$, $B^+_{-\bf q} \equiv B^*$,
$C^+_{\bf q} \equiv C$, $C^+_{-\bf q} \equiv C^*$,
$C^-_{\bf q} \equiv E$, $C^-_{-\bf q} \equiv E^*$.
 Note  that $B^-_{\bf q} = 0$ for the  KH model.
We obtain eight equations which can be written in the matrix form (\ref{m_eq}) with the matrix $\hat{V}$ given by:
\begin{equation}
 {
\small \pmatrix{A&0&C^*&E^*&0&0&B^*&0\cr 0&-A&-E^*&-C^*&0&0&0&-B^*\cr C&E&A&0&B&0&0&0
 \cr -E&-C&0&-A&0&-B&0&0\cr 0&0&-B^*&0&-A&0&-C^*&-E^*\cr 0&0&0&B^*&0&A&E^*&C^*
 \cr -B&0&0&0&-C&-E&-A&0\cr 0&B&0&0&E&C&0&A\cr}.
 }
 \label{A11}
\end{equation}
Substituting the  exchange interaction of the KH model: $J^z_1 = J^z_2 = J$, $J^z_3 = J + K_z$,
$J^+_1 = J + K_x/2$, $J^+_2 = J + K_y/2$, $J^+_3 = J$,
$J^-_1 = K_x/2$, $J^-_2 = -K_y/2$, $J^-_3 = 0$, we obtain  the eigenvalues:
\begin{eqnarray}
&&\varepsilon_{1,2}^2({\bf q}) = A^2 + |C|^2 - |B+E|^2 \pm M_{+},\nonumber\\
&&\varepsilon_{3,4}^2({\bf q}) = A^2 + |C|^2 - |B-E|^2 \pm M_{-},\nonumber\\
&&M^2_{\pm} = 4 A^2|C|^2 - 4 [{\rm Im}(EC^*)]^2 -
4 [{\rm Im}(BC^*)]^2 \nonumber\\
&&\pm 4 [{\rm Re}(B^*\,E^*\,C^2) - |C|^2 \, {\rm Re}(E^*B)],
\label{zz_spec}
\label{A12}
\end{eqnarray}
where
\begin{eqnarray}
&&A = K_z - J, \;\;\; B = J \exp (i{\bf q}\overrightarrow{\delta_3}), \nonumber\\
&&C = (J + K_x/2) \exp (i{\bf q}\overrightarrow{\delta_1}) + (J + K_y/2) \exp (i{\bf q}\overrightarrow{\delta_2}), \nonumber\\
&&E = (K_x/2)\exp (i{\bf q}\overrightarrow{\delta_1}) - (K_y/2)\exp (i{\bf q}\overrightarrow{\delta_2}).
\label{zz_str_e}
\label{A13}
\end{eqnarray}

\subsection{Stripe phase}

Now let us consider the stripe phase (the only difference from the zigzag phase is in the signs of order parameters).
We have the same four sublattices $j = 1,2,3,4$ with the following  signs of the order parameter:
$s(1) = s(4) = 1$, $s(2) = s(3) = -1$.
So we obtain:
\begin{eqnarray}
\gamma^z_1& =& \gamma^z_4 = - \gamma^z_2 =- \gamma^z_3 = A ,
 \nonumber\\
A &=&  J^z_3- J^z_1 - J^z_2,
\end{eqnarray}
\begin{eqnarray}
\gamma^{\pm}_{{\bf q}, 1, 4} = - \gamma^{\pm}_{-{\bf q}, 2, 3}
  = -\gamma^{\pm}_{{\bf q}, 3, 2}=  \gamma^{\pm}_{-{\bf q}, 4, 1}
    = B^{\pm}_{\bf q},&&
\nonumber\\
\gamma^{\pm}_{{\bf q}, 1, 2} = -\gamma^{\pm}_{-{\bf q}, 2, 1}=
- \gamma^{\pm}_{{\bf q}, 3, 4} = \gamma^{\pm}_{-{\bf q}, 4, 3} = C^{\pm}_{\bf q},&&
 \label{A15}
\end{eqnarray}
where the functions $B^{\pm}_{\bf q}$  and $C^{\pm}_{\bf q}$  are given by Eq.~(\ref{A9a}).
By substituting these $\gamma$ functions into Eqs.~(\ref{GF_eq1}), (\ref{GF_eq2}),
with the same A,B,C,E as for the zigzag phase, Eq.~(\ref{zz_str_e}), we obtain the matrix $\hat{V}$ in Eq.~(\ref{m_eq}):
\begin{equation}
{
\small \pmatrix{-A&0&C^*&E^*&0&0&B^*&0\cr 0&A&-E^*&-C^*&0&0&0&-B^*\cr -C&-E&A&0&-B&0&0&0
 \cr E&C&0&-A&0&B&0&0\cr 0&0&-B^*&0&A&0&-C^*&-E^*\cr 0&0&0&B^*&0&-A&E^*&C^*
 \cr B&0&0&0&C&E&-A&0\cr 0&-B&0&0&-E&-C&0&A\cr }.}
\end{equation}
We have  the eigenvalues:
\begin{eqnarray}
&&\varepsilon_{1,2}^2({\bf q}) = A^2 - |C|^2 + |B+E|^2 \pm M_{+}, \nonumber\\
&&\varepsilon_{3,4}^2({\bf q}) = A^2 - |C|^2 + |B-E|^2 \pm M_{-}, \nonumber\\
&&M^2_{\pm}= 4\,A^2\,(|B|^2 + |E|^2) - 4[{\rm Im}(EC^*)]^2 - 4[{\rm Im}(BC^*)]^2 \nonumber\\
&&\pm 4{\rm Re}(B^*\,E^*\,C^2) \pm 4\,(2\,A^2 - |C|^2)\,{\rm Re}(E^*B).
\label{str_spec}
\end{eqnarray}
The equations for $a_{11}$ in the case of the zigzag and stripe phases are too long, so they are computed numerically
by the LU decomposition of a $7\times7$ complex matrix.

\end{document}